\documentclass[12pt,aps]{revtex4}

\begin{document}
\title{Apparent superluminal neutrino propagation caused by nonlinear coherent interactions in matter}

\author{Ram Brustein$^{(1,2,3)}$, Dmitri Semikoz$^{(2,4)}$\\ \  \\
(1)\ Department of Physics,
Ben-Gurion University, Beer-Sheva 84105, Israel \\
(2) APC, 10 rue Alice Domon et Leonie Duquet, F-75205 Paris Cedex 13, France \\
(3) CAS, Ludwig-Maximilians-Universit\"at M\"unchen, 80333 M\"unchen, Germany \\
(4) INR RAS,
 60th October Anniversary pr. 7a, 117312 Moscow,  Russia
\\ email: ramyb@bgu.ac.il, dmitri.semikoz@apc.univ-paris7.fr}

%\maketitle

\begin{abstract}
Quantum coherence can significantly increase the strength of the forward scattering of neutrinos propagating through the Earth and interacting with matter. The index of refraction of the neutrinos propagating in a medium and hence their phase velocity is determined by the forward scattering. So, depending on the nature of the interaction of neutrinos with matter, their phase velocity can be larger than the speed of light in vacuum. We show that such effects can explain the apparent superluminal propagation of muon neutrinos found recently by the OPERA experiment. Our proposal explains why the neutrino oscillations and the propagation of neutrinos from supernova 1987A are unaffected. It can be verified by changing the amount of neutrino coherence or by changing the composition of matter in which they propagate.
\end{abstract}

\maketitle

\section{Introduction}

The OPERA experiment reported~\cite{OPERA_2011} that the time of flight of muon neutrinos propagating through Earth from CERN to the Gran Sasso Laboratory  is relatively shorter by about $2\times 10^{-5}$ than what their time of flight would be if they were propagating at the speed of light in vacuum $c$. Previously, the MINOS experiment has reported a similar measurement with a lower statistical significance \cite{MINOS_2007}. Neutrinos from supernova 1987A were detected on Earth just a few hours before the optical signal~\cite{SN1987a,SN1987b,SN1987c}, which puts a strong constraint on their propagation velocity.

The report of the OPERA experiment prompted us to reconsider the well-known theory of the propagation of neutrinos in matter \cite{wolfenstein} in order to understand its possible relevance. We start by reviewing some essential parts of this theory.

%\section{Index of refraction of neutrinos propagating in matter}

The refraction index in a medium $n$ determines the phase velocity of propagation of (approximately) massless particles through the medium
\begin{equation}
v=\frac{c}{n}.
\label{nu_velocity}
\end{equation}
If $n$ is smaller than unity the phase velocity of the particles through the medium can be larger than $c$. This is, of course, fully compatible with the theory of special relativity.

The wavefunction of $\nu_\mu$'s propagating through a medium is
\begin{equation}
|\nu_\mu\rangle=a e^{\hbox{$i(Et-n\ \vec{k}\cdot\vec{x})$}}.
\label{nu_mu}
\end{equation}
Here we have ignored the interference of $\nu_\mu$ with other flavors of neutrinos and we use units in which $c=1$ and $\hbar=1$.

The refraction index of neutrinos propagating in a medium can be either larger or smaller than unity, depending of the sign of their coherent forward scattering. For muon neutrinos propagating through Earth \cite{wolfenstein},
\begin{equation}
n-1=\sqrt{2} \frac{G_F}{E} \rho_{\rm Earth} \sum\limits_{i=P,N,e} g_i \rho_i
\label{nmin1matter}
\end{equation}
where $G_F$ is the Fermi constant. The number density of nucleons in the Earth $\rho_{\rm Earth}$ is approximately given by the mass density divided by the mass of the proton and $\rho_i$ are the relative number densities of protons, neutrons and electrons. Because matter is neutral $\rho_P=\rho_e$. The couplings $g_i$ can be expressed in terms of the Weinberg angle $\Theta_W$,
\begin{eqnarray}
g_e &=& 2 \sin^2\Theta_W-1/2 \cr
g_P &=& 1/2 - 2 \sin^2\Theta_W \\
g_N &=&-1/2. \nonumber
\end{eqnarray}

Muon neutrinos $\nu_\mu$ (and tau-neutrinos) interact with matter only via a neutral current interaction and it is well-known \cite{wolfenstein} that their index of refraction in matter is smaller than unity. The number density of protons and electrons in the Earth is equal, so the only contribution in this case comes from neutrons. If we assume that the density of neutrons is about equal to that of the protons, then $\sum\limits_{i=P,N,e} g_i \rho_i=-1/4$. The magnitude of the negative deviation of the index of refraction as expressed in Eq.~(\ref{nmin1matter}) for a single neutrino propagating through the Earth is tiny (see details below).

We now arrive at the main point of our paper: that the neutrinos created at the CERN CNGS facility are created in a coherent quantum state with a large intensity and that consequently the magnitude of the negative deviation of the index of refraction is enhanced by a huge factor.

The proton beam in the CERN SPS ring is released on target in a coherent state and the process of extraction of the neutrino beam is executed so as to keep the coherence as much as possible. Then,  when the neutrinos propagate through the Earth they only interact coherently with matter.

Let us describe the production process in more detail.
First, protons are accelerated to an energy of
$E_P= 400$ GeV, focused and tuned with a very small energy spread and spatial cross section. The protons are then extracted and aimed at a graphite target.   Every proton extraction lasts for  10.5 $\mu$s and consist of $2 \times 10^{13}$ protons.  The protons are focused in a beam with spatial cross section of about 0.5 mm \cite{CERN_beam}.

Every proton produces, after hitting the graphite target, some pions. Only part of the pions are useful for the production of the neutrino beam. The useful pions are then focused and collimated and decay in a 1000 m vacuum tunnel into muons and muon neutrinos.

Thus, both the primary protons, the secondary pions and finally the neutrinos  are  very coherent. There are several sources of incoherence in the production line. These include the partial incoherence of the original proton beam, the incoherence of the pion production due to the thermal noise of graphite nuclei and the partial incoherence of the neutrino beam due to finite energy width. We cannot, at the moment, calculate the total amount of decoherence in the production line. We assume in the following that at the end of the production line the neutrino beam is still largely coherent.

If the production process were completely efficient each proton would have produced on average a few pions. In this case every neutrino extraction should have contained ${\rm a\ few} \times 10^{13}$ muon neutrinos.   However, due to various loss factors, each proton produces on average about $0.2$ muon neutrino  and so every extraction contains in total about $4 \times 10^{12}$ muon neutrinos propagating towards the Gran Sasso laboratory. When they reach their destination at a distance of 730 km from their point of origin, the neutrino beam is spread over an area whose effective radius is about 2 km   \cite{neutrino1,neutrino2,neutrino3}. If indeed the amount of decoherence is no too large then the neutrinos travel from CERN to Gran Sasso in coherent waves each consisting a total of about $4\times 10^{12}$ neutrinos.
 
When a coherent wave of neutrinos interacts with matter rather than a single neutrino, then the matter can respond to the wave in a nonlinear way, in analogy with the optical Kerr effect in which the response is proportional to the intensity of the wave. Then the forward scattering amplitude  at zero momentum of the wave  with the matter is enhanced in a way depending quadratically on the amplitude $A$ of the wave or, equivalently, on the number of particles that it consists,
\begin{equation}
(n-1)_{\rm coherent}= b^2 A^2 (n-1),
\label{n_coherent}
\end{equation}
where $b^2$ is a dimensionless parameter that determines the strength of the nonlinear enhancement.  The details of effect depends on how long the medium would ``remember" the propagation of neutrino wave through it. All the macroscopic-size coherent wave can participate in the coherent forward scattering with the same neutron since the forward scattering is at zero momentum, which roughly speaking allows every single neutron to ``see" all the neutrinos of the wave.

We do not know, at the moment, whether an enhancement of the form proposed in Eq.~(\ref{n_coherent}) is possible and if it is possible then how to describe it with a microscopic model. Our attitude is to assume that it is possible and determine the implications of this assumption to the OPERA results.

When reaching Gran Sasso, every coherent neutrino wave interacts with the detector which records its phase:
\begin{equation}
P_\nu=|\langle \nu_\mu | {\rm Detector}  \rangle|^2~,
\end{equation}
where  $ \langle\nu_\mu |$  is neutrino wave function, given by  Eq.~(\ref{nu_mu}) multiplied by the amplitude of the wave $A$.

Thus, the OPERA experiment time shift measurement in effect measures the phase velocity of the coherent neutrino wave given by  Eq.~(\ref{nu_velocity}) which depends on the refraction index given by  Eq.~(\ref{n_coherent}). The index of refraction is smaller than unity and so the phase velocity is larger than $c$ and neutrinos appear to arrive too early. It is possible to verify that the group velocity $v_g=dE/dk$ in this case remains equal of $c$ \footnote{Up to (negative) corrections that are second order in the small parameter $(n-1)_{\rm coherent}$.}, as it should be for (approximately) massless neutrinos. Additionally, the velocity of propagation of the leading front of the wave is in our case equal to $c$ both because it is related to the group velocity and because we do not expect a significant nonlinear enhancement  of $n-1$ for it.  In any case, it is unclear that the OPERA experiment is measuring the speed of propagation of a signal (or information) from CERN to Gran Sasso.

Let us turn to an order of magnitude estimate of the effects that we have just discussed.
The refraction index for a coherent neutrino wave can be reexpressed using Eq.~(\ref{nmin1matter}) and Eq.~(\ref{n_coherent}) as
\begin{equation}
(n-1)_{\rm coherent}\simeq - 2.5\times 10^{-5} \left(\frac{\rho_{\rm Earth}}{3 g/cm^3}\right)\left(\frac{17 GeV}{E}\right)\left(\frac{b A}{ 2 \times 10^{9}}\right)^2.
\label{nmin1quant}
\end{equation}
Here we have used $\sqrt{2} G_F=(1/246\ {\rm GeV})^2$ and used for normalization typical values of rock density in the Earth and neutrino energy. 

The degree of coherence and nonlinear enhancement that will be needed to explain the OPERA result can be read off Eq.~(\ref{nmin1quant}). It requires that $b A \sim 2\times 10^{9}$. We would like to point out that this is exactly the number of neutrinos in each of the 2000 bunches contained in every extraction. 

Based on the considerations that we have described so far, we propose that OPERA experiment measured the phase velocity of a coherent neutrino beam propagating in the Earth.  We conclude that all the results presented in \cite{OPERA_2011} can be made consistent with the theory of special relativity, quantum mechanics and the Standard Model of particle interactions, provided that the produced neutrino beam is coherent enough and a that a large enough amount of nonlinear enhancement takes place. 

The MINOS experiment uses the NuMI beam which contains about $3\times 10^{13}$ protons per bunch \cite{MINOS_beam}, so we expect that the degree of coherence that is required to explain the MINOS result is similar to that required to explain the OPERA results. 

Our proposed explanation for the origin of the superluminal neutrino propagation detected by OPERA  is consistent with the data about neutrinos from supernova  SN 1987A. The proposed effect depends on having a coherent neutrino wave interacting with matter which does apply in the case of propagation of  neutrinos from supernovae.

Also, the effect that we have described does not influence the propagation and oscillations of solar neutrinos, atmospheric neutrinos that will not be affected at all by the coherent enhancement that we have described. The considerations about neutrino oscillations in neutrinos produced in coherent waves will also be affected in a negligible way by the enhancement of the forward scattering amplitude at zero momentum.

Our explanation can be verified by verifying its two basic ingredients: that the effect is due to interactions of neutrinos with matter and that it is due to the coherent nature of the neutrino wave. The amount of coherence of the wave can be modified by modifying the properties of the proton beam from which it is produced   such as the number of protons in a bunch,  or the amount of the bunch quantum squeezing. One could also modify the properties of the medium by having the neutrinos propagate through the core of the Earth or through air. In each case, Eq.(\ref{nmin1quant}) predicts a specific dependence on these modifications.

Our treatment emphasizes the coherent nature of neutrino production and propagation in long baseline experiments and suggests that such experiments can be used to test many fundamental aspects of quantum mechanics on scales of thousands of kilometers with high precision.

\acknowledgments
We would like to thank Pierre Binetruy,   Stavros Katsanevas and David Langlois for helpful discussions. We are grateful to Nikos Vassilopoulos for providing us information on the CNGS neutrino beam.  Finally, we thank Alexander Dolgov and Shmuel Nussinov for critical reading of the manuscript and useful comments.
The research of RB is supported  by Israel Science Foundation grant 239/10.

\end{document}